# Manipulation and steering of hyperbolic surface polaritons in hexagonal boron nitride

S. Dai[1,2], M. Tymchenko[2], Y. Yang[3], Q. Ma[3], M. Pita-Vidal[3], K. Watanabe[4], T. Taniguchi[4], P. Jarillo-Herrero[3], M. M. Fogler[1], A. Alù[2]*, D. N. Basov[1,5]*

[1]*Department of Physics, University of California, San Diego, La Jolla, California 92093, USA*

[2]*Department of Electrical & Computer Engineering, The University of Texas at Austin, Texas 78712, USA*

[3]*Department of Physics, Massachusetts Institute of Technology, Cambridge, Massachusetts 02215, USA*

[4]*National Institute for Materials Science, Namiki 1-1, Tsukuba, Ibaraki 305-0044, Japan*

[5]*Department of Physics, Columbia University, New York, New York 10027, USA*

*Correspondence to: db3056@columbia.edu, alu@mail.utexas.edu

Abstract:

Hexagonal boron nitride (hBN) is a natural hyperbolic material that supports both volume-confined hyperbolic polaritons (HPs) and sidewall-confined hyperbolic surface polaritons (HSPs). In this work, we demonstrate effective excitation, control and steering of HSPs in hBN through engineering the geometry and orientation of hBN sidewalls. By combining infrared (IR) nano-imaging and numerical simulations, we investigate the reflection, transmission and scattering of HSPs at the hBN corners with various apex angles. We show that the sidewall-confined nature of HSPs enables a high degree of control over their propagation by designing the geometry of hBN nanostructures.

Hyperbolic materials are anisotropic media for which the principal components of the permittivity tensor have opposite signs[1]. This class of natural and artificial materials has led to a series of major advances in nanophotonics, including but not limited to sub-diffraction imaging[2, 3], negative refraction[4], and emission energy control[5]. Metamaterials[1, 3, 4, 6, 7] and metasurfaces[8-10] comprised of artificially fabricated/stacked nanostructures are most commonly employed to realize either three-dimensional (3D) or two-dimensional (2D) hyperbolic dispersions. In parallel with advances in metamaterials, 3D and 2D hyperbolicity has been theoretically predicted and recently discovered experimentally in a range of natural materials[11-22]. Many natural hyperbolic materials belong to the class of van der Waals (vdW) crystals[23], highly anisotropic compounds in which atomic layers are bonded together through vdW forces[24]. Hexagonal boron nitride (hBN) is a prototypical vdW material naturally possessing both Type I ($\varepsilon_{xy} > 0$, $\varepsilon_z < 0$, $\varepsilon_{xy}$ and $\varepsilon_z$ are permittivity in the basal plane and along the optical axis) and Type II ($\varepsilon_{xy} < 0$, $\varepsilon_z > 0$) 3D hyperbolicity that stems from its anisotropic phonon resonances[11, 12, 14, 25, 26]. Recent works have reported on the nano-imaging of propagating polaritons – electromagnetic waves hybridized between photons and electromagnetic dipole modes[23, 27, 28] – in hyperbolic systems. In hBN, both volumetric hyperbolic polaritons (HPs)[11, 14, 15, 25, 29-31] and surface-confined hyperbolic surface polaritons (HSPs)[20] have been observed. HPs and HSPs have been shown to possess significantly improved polaritonic loss and upper momentum cut-off due to the hBN's natural lattice[29, 30]. The latter is superior to those of metamaterials and metasurfaces which inevitably suffer from finite feature sizes of constituent elements and fabrication imperfections. While previous works focused on the intrinsic properties of HSPs[17, 20], here we combine infrared (IR) nano-imaging and finite element simulations to demonstrate the manipulation of reflection, transmission, scattering and propagation steering of HSPs in tailored hBN nanostructures (see Methods).

Infrared nano-imaging experiments were performed in the Type II hyperbolic region of hBN using scattering-type scanning near-field optical microscopy (s-SNOM, methods)[11, 14, 25, 30]. A metalized atomic force microscope (AFM) tip was illuminated by infrared (IR) light (Fig. 1a), generating strong electromagnetic near fields underneath the apex. These fields excite a wide range of momenta $q$ over the surface, and therefore facilitate the formation of coupled photon-lattice modes referred to as phonon polaritons. In the experiment, the AFM tip acts both as a launcher[32] and a detector of polaritons supported in hBN. Standing wave oscillations of the propagating polaritons are recorded as samples are scanned underneath the AFM tip[11, 14, 25, 30]. The s-SNOM images of hBN nanostructures at a representative IR frequency $\omega = 1425$ cm$^{-1}$ ($\omega = 1/\lambda_{IR}$ where $\lambda_{IR}$ is the IR wavelength) are presented in Figure 1. On the top surface of a hBN cuboid (Fig. 1a), we observe a series of oscillations of nano-IR signal that we represent in the form of scattering amplitude $s(\omega)$. In Fig. 1, one can recognize bright fringes close to the sidewalls, and slightly damped interference hot spots in the interior of the top surface. As established in previous works[11, 14, 25, 30], these features originate from the standing wave interference between HPs launched by the AFM tip and those reflected by the hBN sidewalls. The HPs propagate as guided waves inside the hBN slab, undergoing multiple reflections at the top and bottom surfaces of the slab[29]. The period of polariton fringes on the hBN top surface equals to one half of the HP wavelength $\lambda_{HP}/2$.

In addition to the hallmarks of HPs, another series of polaritonic fringes appears along the sidewalls of the hBN slab. These latter fringes exhibit the strongest amplitude close to the hBN sidewall corner followed by slightly damped features away from the corner (Fig. 1a). While these fringes display qualitatively similar interference patterns with those of the HPs, the sidewall polaritons reveal distinct oscillation periods. In Figure 2a, we plot the s-SNOM line traces taken along the red dashed and solid line cuts in Figure 1a. The polaritonic fringes in Figure 1a appear as oscillation peaks in the s-SNOM line traces (Figure 2a). Compared with the HPs (red dashed line), polaritons along the sidewall (red solid line) possess an oscillation period ~ 20% shorter. Therefore, these polaritons can be regarded as HSPs confined along the hBN slab sidewalls, and undergo multiple reflections between the top and bottom edges when propagating therein[20]. The formation of these propagating HSPs can be attributed to the 2D Type II hyperbolicity of the hBN sidewall[10, 17, 20]. The technique used to image HSPs is akin to that of HPs, except the reflectors of propagating polaritons (to form polariton standing wave fringes) are slab corners instead of slab sidewalls in hBN. The period of the polariton fringes on the sidewalls therefore equals to $\lambda_{HSP}/2$. Similar interference patterns originating from edge plasmon polaritons in graphene nanostructures have also been reported recently[33, 34].

Having established polaritonic edge modes, we now explore their propagation along the perimeter of diamond-shaped hBN flakes, and also along the circumference of hBN disks (Figs. 1a, b, c and d). Evident HSP fringes are observed around the hBN sidewall corners with an angle $\alpha$ in the range of 90º to 180º. Additional s-SNOM images of the hBN slabs are provided in Supplementary Section 1. We observe bright spots close to the hBN corner with $\alpha < 90º$ (Fig. 1c-d). However, we do not attribute these features to polaritonic fringes in the following analysis, since the periodic oscillation fringes are missing. The amplitude of the HSP fringe oscillation decreases with increasing angle (Figs. 1a, c, d and Fig. 2a), and vanishes in the hBN disk (Fig. 1b) where no evident HSP features can be observed.

The evolution of HSP fringes as a function of the hBN corner angle reveals the possibility of controlling and manipulating the propagation of HSPs by tailoring the hBN corner angle. HSPs launched with the s-SNOM tip propagate along the sidewall (green arrow in Figure 2c) and can be reflected, transmitted and/or scattered into volume-confined HPs when reaching the corner. To inquire into the laws governing HSP propagation, we first performed finite element simulations of HSPs at hBN flake's corners with various apex angles α, using COMSOL Multiphysics. Simulations with representative corner angle α = 125º and 300º are shown in Figures 2c and 2d. In our simulations, the HSPs are launched from Port 1 (yellow arrow), and then propagate along the sidewall I towards the corner. Note that we have filtered volume confined HPs simultaneously launched from Port 1 in order to isolate the properties of HSPs. Upon reaching the corner, the propagating HSPs have three principal options: 1) back reflection towards Port 1 along sidewall I (red arrow); 2) transmission through the corner and propagation towards Port 2 along sidewall II (cyan arrow); 3) scattering at the corner with conversion into volume confined HPs towards the interior of the hBN slab (grey arrows). Propagating HSPs (yellow arrow) may also get scattered into free-space IR photons at the corner, however in our simulation this latter channel produced only negligible corrections, because of the tight confinement of HSPs to the surface and large momentum mismatch with free-space photon. The reflection $R$, transmission $T$ and scattering $S$ coefficients at the hBN corner can be defined as the ratio of the electric field of polaritons in each channel over that of the incident HSPs (yellow arrow): $X = E_x / E_{in}$ ($X = R$, $T$ or $S$, respectively). By exploring the polariton energy in different ports, we can determine the fraction of HSPs reflected ($R$), transmitted ($T$) and scattered into volume-confined HPs ($S$) as a function of the corner angle α (Fig. 2b). The reflection coefficient $R$ (red squares in Fig. 2b) can be obtained by measuring the standing wave oscillation intensity of the HSP fringes formed in the round-trip imaging mechanism[35]. The simulation results agree excellently with our experimental data.

The polariton energy partition among these three channels (characterized by coefficients $R$, $T$ and $S$) varies dramatically, with the corner angle α (Fig. 2b). At α ≈ 95º, nearly all incident HSPs are back-reflected towards the Port 1 whereas the transmission $T$ and scattering $S$ are negligible.

The reflection $R$ decreases while transmission $T$ and scattering $S$ grow with the increasing $\alpha$. At $\alpha \sim 125°$, the scattering coefficient $S$ becomes comparable to $R$ and $T$, indicating a considerable portion of HSPs scattered at the corner and converted into HPs in the hBN interior (see also Fig. 2c). At $\alpha = 180°$, $R$ and $S$ both vanish such that all the HSPs get transmitted, as expected in the absence of a corner. As $\alpha$ is further increased $R$ and $T$ reveal the opposite trends whereas the scattering of HSPs ($S$) increases significantly and exceeds 60% at the angle $\alpha = 300°$. In the latter regime, HPs scattered from the HSPs can be clearly observed (Fig. 2d).

Similar to published demonstrations for HPs in hBN[11, 12, 14], the manipulation of HSPs can also be accomplished by changing the hBN slab thickness: a consequence of the waveguiding nature of propagating HSPs. The hBN thickness can be easily altered via mechanical exfoliation, since the hBN layers are bonded together via weak van der Waals forces[24]. At the representative IR frequency $\omega = 1425$ cm$^{-1}$, we demonstrate manipulation of the HSP wavelength by varying the thickness of the hBN slab (Fig. 3). The HSP wavelength scales with the hBN thickness following a nearly linear law (red squares and solid line for data and simulation, respectively), consistent with the waveguided nature of HSPs. At this frequency, the HPs also obey a similar scaling law (red dots and dashed line), as reported before[11, 12, 14].

We finally highlight the potential of steering HSPs by engineering the geometry of an hBN slab (Figure 4). Propagation steering of HPs and other polaritonic modes seem arduous without detailed design of the local dielectric environment in sophisticated structures[36-38], since these polaritons can propagate along any direction in the basal plane ($k_{//,HP}$, blue arrows) once they are launched. The surface-confined nature of HSPs restricts the propagation ($k_{//,HSP}$, red arrows) along any chosen direction of the hBN sidewalls. In a proof-of-concept implementation, we engineered the hBN structure in Figure 4, observing HSP fringes not only along the straight sidewalls but also along the circumference of the fabricated semicircles. The propagation trace of HSPs can switch between a straight line and a designed curve (red arrows), following the engineered hBN sidewall orientation. Through propagation steering of polaritons, HSPs can travel around the semi-circle region (region A). Based on this concept, we envision the possibility of designing proper phase accumulation to realize a polaritonic cloak[39, 40]: an important element of transformation polaritonics[36, 37].

The results presented in Figures 1 to 4 demonstrate the manipulation and steering of HSPs in hBN. The sidewall-confined nature of HSPs enables an efficient control of polariton reflection, transmission, scattering and propagation through tailoring the geometry of hBN nanostructures. The methodology that we utilized to manipulate and steer HSPs in hBN can be directly extended to surface polaritons propagating in hyperbolic metamaterials, metasurfaces[10, 16] and other van der Waals materials, including black phosphorus[13] and topological insulators[41]. We envision future efforts directed towards hybridizing[42, 43] HSPs with other polaritonic modes, such as plasmons in graphene and black phosphorus, for the realization of dynamically tunable hyperbolic metasurfaces and polaritonic flat optics[16, 38, 44]. The propagation steering demonstrated in Fig. 4 may provide advantages for the implementation of transformation optics/polaritonics[36, 37] over alternative platforms, by employing polaritons propagating along the sidewall of natural or artificial hyperbolic materials to engineer advanced waveguiding functionalities.

**Methods**

*Experimental setup*

The infrared (IR) nano-imaging experiments introduced in the main text were performed using a scattering-type scanning near-field optical microscope (s-SNOM). Our s-SNOM is a commercial system (www.neaspec.com) based on a tapping-mode atomic force microscope

(AFM). In the experiments, we use a commercial AFM tip (tip radius ~ 10 nm) with a $PtIr_5$ coating. The AFM tip is illuminated by monochromatic quantum cascade lasers (QCLs) ([www.daylightsolutions.com](www.daylightsolutions.com)) covering a frequency range of 900 – 2300 $cm^{-1}$ in the mid-IR. The s-SNOM nano-images were recorded by a pseudo-heterodyne interferometric detection module with an AFM tapping frequency 280 kHz and tapping amplitude around 70 nm. In order to subtract background signal, the s-SNOM output signal was demodulated at the $3^{rd}$ harmonics of the tapping frequency.

*Sample fabrication*

Hexagonal boron nitride (h-BN) crystals were mechanically exfoliated from bulk samples and deposited onto Si wafers capped with 285 nm thick $SiO_2$. Oxygen plasma etching was ultilized to pattern the hBN slabs into anticipated geometries.

**Author contributions:**

All authors were involved in designing the research, performing the research and writing the paper.

**Competing interests:**

None.

**Acknowledgments:**

Work at UCSD and Columbia on optical phenomena in vdW materials is supported by the Gordon and Betty Moore Foundation's EPiQS Initiative through Grant GBMF4533. Research at UCSD and Columbia on metamaterials and development of nano-IR instrumentation is supported by AFOSR FA9550-15-1-0478, and ONR N00014-15-1-2671. P.J-H acknowledges support from AFOSR grant number FA9550-11-1-0225. Research at UT Austin has been supported by the AFOSR MURI grant number FA9550-17-1-0002.

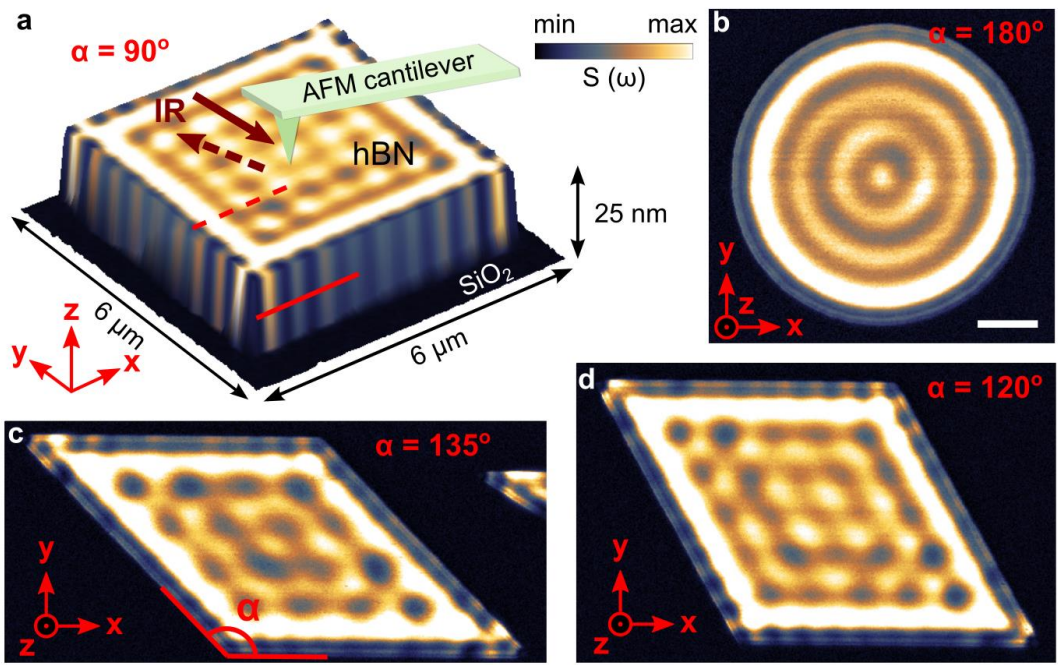

**Figure 1 | Hyperbolic surface polaritons in hBN nano diamonds. a**, Experiment setup. The AFM tip is illuminated by an IR beam from QCL. **b – d,** s-SNOM images of HSPs in hBN nano diamonds with corner angle 180º (**b**), 135º (**c**) and 120º (**d**) at ω = 1425 cm$^{-1}$. Thickness of the hBN: 25 nm. Scale bar: 1 μm.

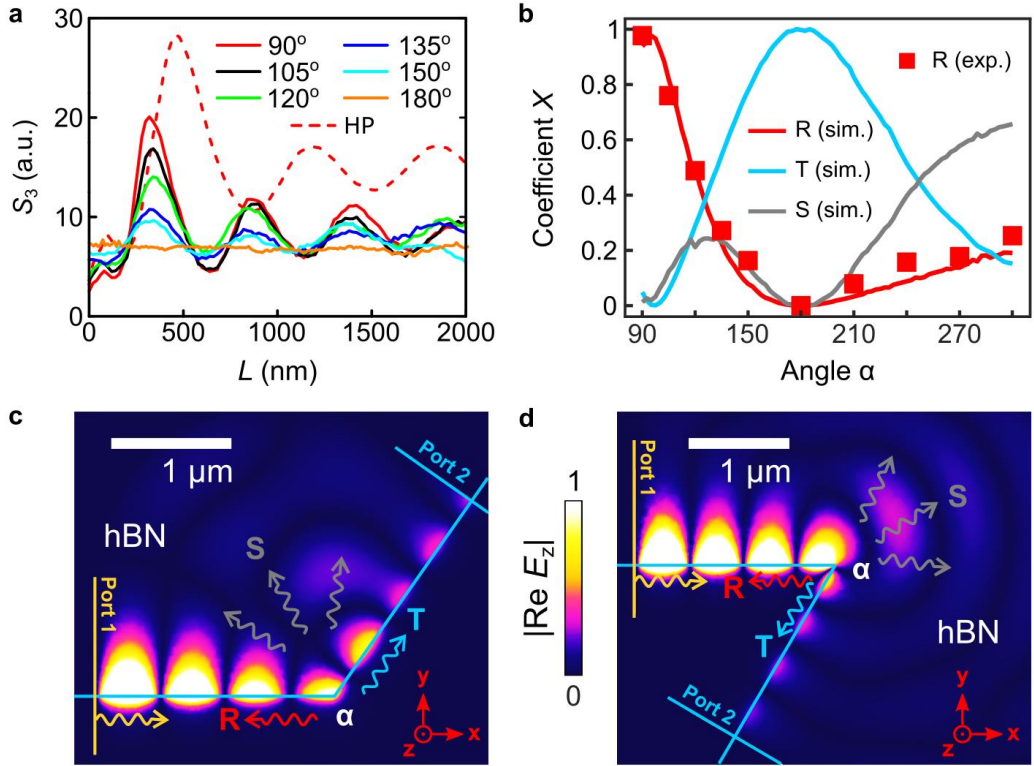

**Figure 2 | Reflection, transmission and scattering of HSPs at the hBN corners. a,** s-SNOM line profiles of HSPs at hBN slab corners with various apex angles. **b,** Simulated reflected (R), transmitted (T) and scattered to HPs (S) fractions of HSP at 25nm-thick hBN flake's corner as a function of the apex angle $\alpha$. Red squares indicate the experimental data of the HSP reflection extracted from s-SNOM images at IR frequency $\omega$ = 1425 cm$^{-1}$. **c, d,** Simulated spatial distributions of $|\text{Re}\,E_z|$ of the fundamental HSPs mode as it reaches the hBN flake's corners with apex angles $\alpha$ = 127º (**c**) and $\alpha$ = 300º (**d**).

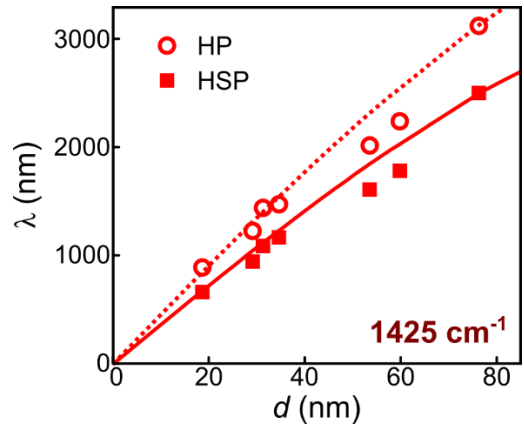

**Figure 3 | Thickness dependence of HSP and HP in hBN.** Dots and squares, experimental data extracted from s-SNOM images for HPs and HSPs in hBN, respectively. IR frequency $\omega = 1425$ cm$^{-1}$.

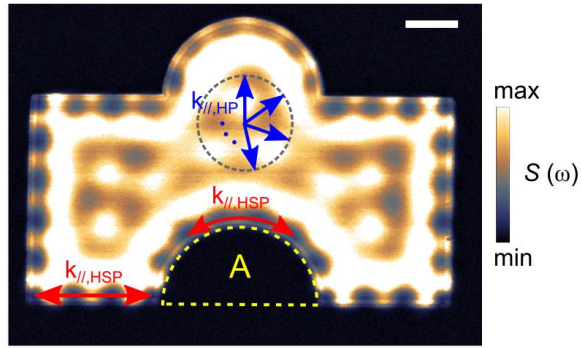

**Figure 4 | Propagation steering of HSPs in hBN.** s-SNOM image of HSPs in hBN with engineered geometry at $\omega = 1410$ cm$^{-1}$. Blue and red arrow indicate the propagation direction of HPs and HSPs in hBN, respectively. Scale bar: 1 μm.